\documentclass[
aps,%
11pt,%
final,%
notitlepage,%
oneside,%
twocolumn,%
nobibnotes,%
nofootinbib,%
superscriptaddress,%
noshowpacs,%
centertags]%
{revtex4}
\usepackage[koi8-r]{inputenc}
\usepackage{multirow}
\begin{document}
\selectlanguage{english}

\title{Study of Envelope Velocity Evolution of Type
Ib-c Core-Collapse Supernovae from Observations of  XRF 080109 / SN 2008D and
\linebreak  GRB 060218 / SN 2006aj with BTA}

\author{\firstname{A.~S.}~\surname{Moskvitin}}
\affiliation{\saoname}

\author{\firstname{E.}~\surname{Sonbas}}
\affiliation{University of Adiyaman, Department of Physics, 02040
Adiyaman, Turkey}

\author{\firstname{V.~V.}~\surname{Sokolov}}
\affiliation{\saoname}

\author{\firstname{T.~A.}~\surname{Fatkhullin}}
\affiliation{\saoname}

\author{\firstname{A.~J.}~\surname{Castro-Tirado}}
\affiliation{Instituto de Astrof\'{\i}sica de Andaluc\'{\i}a
(IAA-CSIC), P.O. Box 03004, 18080 Granada, Spain}

\begin{abstract}
Results of modeling the spectra of two supernovae SN\,2008D and
SN\,2006aj related to the \linebreak X-ray flash XRF\,080109 and
gamma-ray burst GRB\,/\,XRF\,060218, respectively, are studied.
The spectra were obtained with the 6-meter BTA telescope of the
Special Astrophysical Observatory of the Russian Academy of
Sciences in 6.48 and 27.61 days after the explosion of SN\,2008D,
and in 2.55 and 3.55 days after the explosion of SN\,2006aj. The
spectra were interpreted in the Sobolev approximation with the
\texttt{SYNOW} code. An assumption about the presence of envelopes
around the progenitor stars is confirmed by an agreement between
the velocities of lines interpreted as hydrogen and helium, and
the empiric power-law velocity drop with time for the envelopes of
classic core-collapse supernovae. Detection of a P Cyg profile of
the H$\beta$ line in the spectra of optical afterglows of GRBs can
be a determinative argument in favor of \mbox{this hypothesis.}
\end{abstract}
\maketitle
\section{INTRODUCTION}

The modern classification of core-collapse supernovae includes
several subtypes. If the spectra demonstrate hydrogen lines, then
the supernova belongs to the II type. Its light curve shape is
determined by the hydrogen recombination features in the
envelopes. If a star loses its hydrogen envelope, but the helium
shell is still present in the spectrum, then the star is
classified into the Ib type. The impacts of hydrogen recombination
are absent in its light curve, and its shape is determined by the
heating of the emitting matter as a result of the \linebreak
$^{56}{\rm Ni} \rightarrow ~^{56}{\rm Co} \rightarrow ~^{56}{\rm
Fe}$ decay, depending on the explosion energy. If the helium
envelope is lost and the spectra show only the traces of heavier
elements \mbox{(O, Mg, Si, S, Ca, Fe),} then the supernova is
classified into the Ic type. Different subtypes of supernovae vary
by different explosion energies, and, consequently, by the
velocities of matter expansion in the envelopes.

Since in some cases the classification depends on the phase of the
supernova at the moment of observation, some intermediate types
are introduced: Ib/c (appearance of weak helium lines in late
spectra), IIb (disappearance of hydrogen lines in late spectra).
The classification of core-collapse supernovae can be conceived as
a manifold of intermediate subtypes, depending continuously on the
initial parameters of the progenitor star.

%some events show ambiguous behavior at different stages of
%evolution forming intermediate classes Ib/c (appearance of weak
%helium lines in later spectra), IIb (disappearance of helium lines
%in later spectra), the supernovae classification could be
%conceived with more diffuse, smooth borders depending continuously
%on initial parameters of a progenitor star.

This paper discusses the manifestations of stellar-wind envelopes
around the progenitor stars of an \mbox{X-ray} flash XRF\,080109
and a gamma-ray burst with a strong X-ray component GRB /
XRF\,060218. Some signs of supernovae (SNe) were noticed in the
spectra and light curves of both events, which allowed studying
these phenomena from the very start of the explosion (unlike most
SNe that are detected at the moments close to the ``classical''
maximum of the light curve). The study of this kind of phenomena
makes it possible to approach the solution to the supernova
explosion mechanism and the origin of gamma-ray bursts.

On January 9.57, 2009 (UT) the X-ray telescope XRT on board of the
\textit{Swift} space observatory registered an X-ray flash
XRF\,080109. The X-ray component was detectable for about 15
\mbox{minutes \cite{Chevalier_Fransson2008:Moskvitin_n,
Soderberg2008:Moskvitin_n}.} The Burst Alert Telescope (BAT) did
not detect any gamma-ray quanta. The passage of a shock wave
through the envelope was observed in multi-color light curves
during several days \cite{Soderberg2008:Moskvitin_n,
Modjaz2008:Moskvitin_n}. The afterglow light curves and spectra
distinctly showed signs of the supernova, designated as SN\,2008D.

Almost two years earlier (on February 18.149, 2006) the same space
platform \textit{Swift} detected a peculiar gamma-ray burst with
the SN signs in the afterglow spectra and light curves of
GRB\,/\,XRF\,060218\,/ \,SN\,2006aj. It is the first phenomenon
for which we observed the passage of a shock wave through the wind
envelope surrounding a massive SN progenitor star, which
manifested itself as a thermal component in the X-ray spectrum
observed during the first 2 hours, and then as a powerful UV burst
with the maximum observed in 11 hours after the burst
~\mbox{\cite{Blustin2007:Moskvitin_n, Campana2006:Moskvitin_n}}.
The effect of the shock wave passage is seen in the optical light
curve as a relatively short peak several days of duration
\cite{Campana2006:Moskvitin_n, Sonbas2008:Moskvitin_n}.

%(see Fig.2 in \cite{Campana2006} and Fig.2 in our previous paper
%\cite{Sonbas2008}).

In both cases the moment of the X-ray flash and/or gamma-ray burst
registration was taken as the beginning of the SN explosion.

The structure of the paper is as follows: spectral observations
and preliminary data processing are described in section
\ref{SpecObs:Moskvitin_n}; the interpretation of the obtained
spectra with the \texttt{SYNOW} code is laid out in
\ref{SYNOW:Moskvitin_n}; a comparison of the obtained measurements
of the photosphere and envelope expansion velocities with the
analogous velocities of SNe that did not show any relation with
gamma-ray bursts or X-ray flashes is in section
\ref{Vel_phot_env:Moskvitin_n}; our conclusions are discussed in
section \ref{results:Moskvitin_n}.

\section{SPECTRAL OBSERVATIONS}\label{SpecObs:Moskvitin_n}

Optical afterglows of XRF\,080109 and GRB\,/ \,XRF\,060218 were
observed with the \textit{SCORPIO} Optical Reducer mounted in the
primary focus of the 6-m BTA telescope of the Special
Astrophysical Observatory of the Russian Academy of Sciences (SAO
RAS). The VPHG550G grism covering the spectral range of
3500--7500\,\AA{} with a resolution (FWHM) of 10\,\AA{} was used
as a dispersing element.

We obtained two spectra of SN\,2008D: on January 16 and February
6, 2008 (6.48 and 27.61 days after the explosion). For SN\,2006aj
two spectra were obtained as well: on February 20 and 21, 2006
(2.55 and 3.55 days after the explosion). The data processing was
standard and included: subtraction of the electronic zero (an
additive component of the total signal produced by the CCD chip),
flat-field correction (compensation of the CCD chip sensitivity
irregularities), wavelength calibration with the aid of a
comparison spectrum of a Ne--Ar lamp, atmosphere extinction
correction, and calibration by the absolute flux with the use of
observations of a spectrophotometric standard performed every
night. Apart from that, the spectra of SN\,2008D were subtracted
for the contribution of the host galaxy, the spectral distribution
of which is constructed from the regions located in immediate
proximity to the supernova.

Then the observed spectra were corrected for the extinction in the
Galaxy according to the dust distribution maps
\cite{Schlegel1998:Moskvitin_n}. When taking the extinction into
account, the dust-screen model was accepted, where the expression
to account for extinction is of the form \mbox{$F_{int}(\lambda) =
F_{obs}(\lambda)10^{0.4\times k(\lambda)E(B-V)}$}, where
$F_{int}(\lambda)$ and $F_{obs}(\lambda)$ are the emitted (without
absorption) and observed fluxes, respectively. The Milky Way
extinction curve $k(\lambda)$ was taken from the paper
\cite{Cardelli1989:Moskvitin_n}. The spectra of SN\,2008D were as
well corrected for the extinction in the host galaxy according to
the data from \cite{Modjaz2008:Moskvitin_n}. The absorption in the
host galaxy of SN\,2006aj is negligible
\cite{Sollerman2006:Moskvitin_n}.

\begin{figure*}
\setcaptionmargin{5mm}
\onelinecaptionsfalse
    \includegraphics[width=15cm, bb=20 10 800 500, clip]{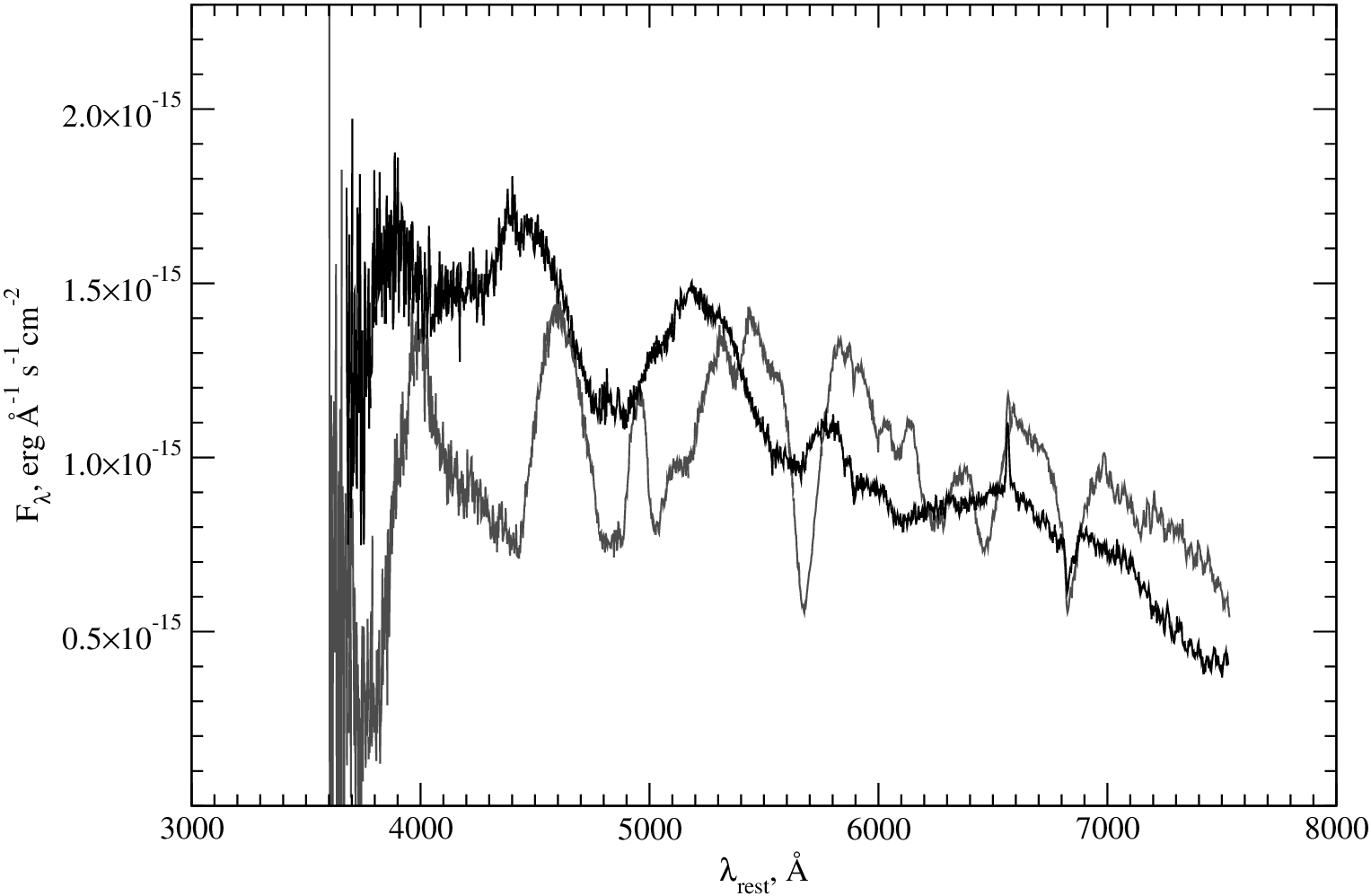}
   \captionstyle{normal}
\caption{Spectra of the supernova SN\,2008D obtained with the BTA
on January 16 (black line) and February 6 (grey line), 2008 in
6.48 and 27.62 days after the explosion, respectively. }
\label{two:Moskvitin_n}
\end{figure*}

Before the interpretation,  the spectra were transferred to the
reference frames, associated with the gamma-ray burst or the X-ray
flash (\mbox{$z=0.0331$} for GRB\,/\,XRF\,060218 and
\mbox{$z=0.007$} for XRF\,080109, see Fig. \ref{two:Moskvitin_n}).
The redshifts were estimated from the shift of galactic emission
lines, and are consistent with data published in the literature.

%The observed spectra (see Fig.\ref{two:Moskvitin_n}) vary by the
%spectrum slope in the continuum , as well as by the depth and shift of
%the spectral features. A decrease in the inclination of the
%spectrum indicates a decrease of temperature, describing the
%energy distribution in the continuum. An increase in the
%absorption depth shows an increase of opacity in the lines, and
%the shift to the red side of their minima is associated with a
%decrease of the observed expansion velocities.

%   The program of the special observations of the optical afterglow with the 6-m telescope included the
%   spectroscopy of the variable object discovered by $\textit{Swift}$ observatory and identified with
%   both XRF\,060218 and XRF\,080109. We used the \textit{SCORPIO} focal reducer mounted in the primary focus of the
%   6-m telescope. The XRF\,060218 observations were made on February 20, 2006 and February 21, 2006.
%   XRF\,080109 was observed on January 16, 2008 and February 6, 2008. As the despersive element we used
%   a VPHG550G grism with an operating wavelength interval
%   3500 - 7500 $\AA$ and resolution (FWHM) of 10 $\AA$.
% < SN 2006aj >
%
% < SN 2008D >
%

\section{COMPARISON OF THE OBSERVED AND SYNTHETIC SPECTRA}\label{SYNOW:Moskvitin_n}

To interpret the observed spectra in detail, we applied the
multi-parametric \texttt{SYNOW} code
\cite{Branch2001:Moskvitin_n}, which was previously used for the
analysis of spectra of core-collapse
supernovae~\mbox{\cite{Branch2002:Moskvitin_n,
Baron2005:Moskvitin_n, Elmhamdi2006:Moskvitin_n}}. The code
algorithm is based on the following assumptions: spherical
symmetry; homologous expansion of \mbox{layers ($v \sim r$);}
sharp border of the photosphere emitting a black-body spectrum and
%%%%%%%%%%%%%%%%%%%%%%%%%%%%%%%%%%%%%%%%%%%%%%%%%%identified
associated at early stages with a shock wave.

The code is used to identify the lines and find the expansion
velocities of the layers in which they are formed. The P Cyg line
profiles observed in the spectra of supernovae and modeled with
the code are then divided into two types according to their shape:
\begin{itemize}
 \item the lines formed in the layer undetached from the expanding photosphere;
  %the undetached case --- a layer where the line is formed is not detached
  %from the expanding photosphere;
 \item the lines appearing in the layer detached from the photosphere.
%the detached case --- a layer is detached from the photosphere.
\end{itemize}

The version of undetached layers was used for the first spectrum
of SN\,2006aj. A combined version in which lighter ions are
detached and the heavy ions are not detached from the photosphere
was applied for the second spectrum of SN\,2006aj and for both
spectra of SN\,2008D. The version choice was determined by fitting
the parameters of every ion to the observed spectral features.

The model parameters of the input file and examples of
interpretation of the SNe spectra are described in detail in the
papers by code developers \cite{Branch2001:Moskvitin_n,
Branch2002:Moskvitin_n, Baron2005:Moskvitin_n,
Elmhamdi2006:Moskvitin_n}.

\subsection{Modeling of SN\,2008D Spectra}

\begin{figure*}
\setcaptionmargin{5mm}
\onelinecaptionsfalse
   \includegraphics[width=15cm, bb=20 10 800 500, clip]{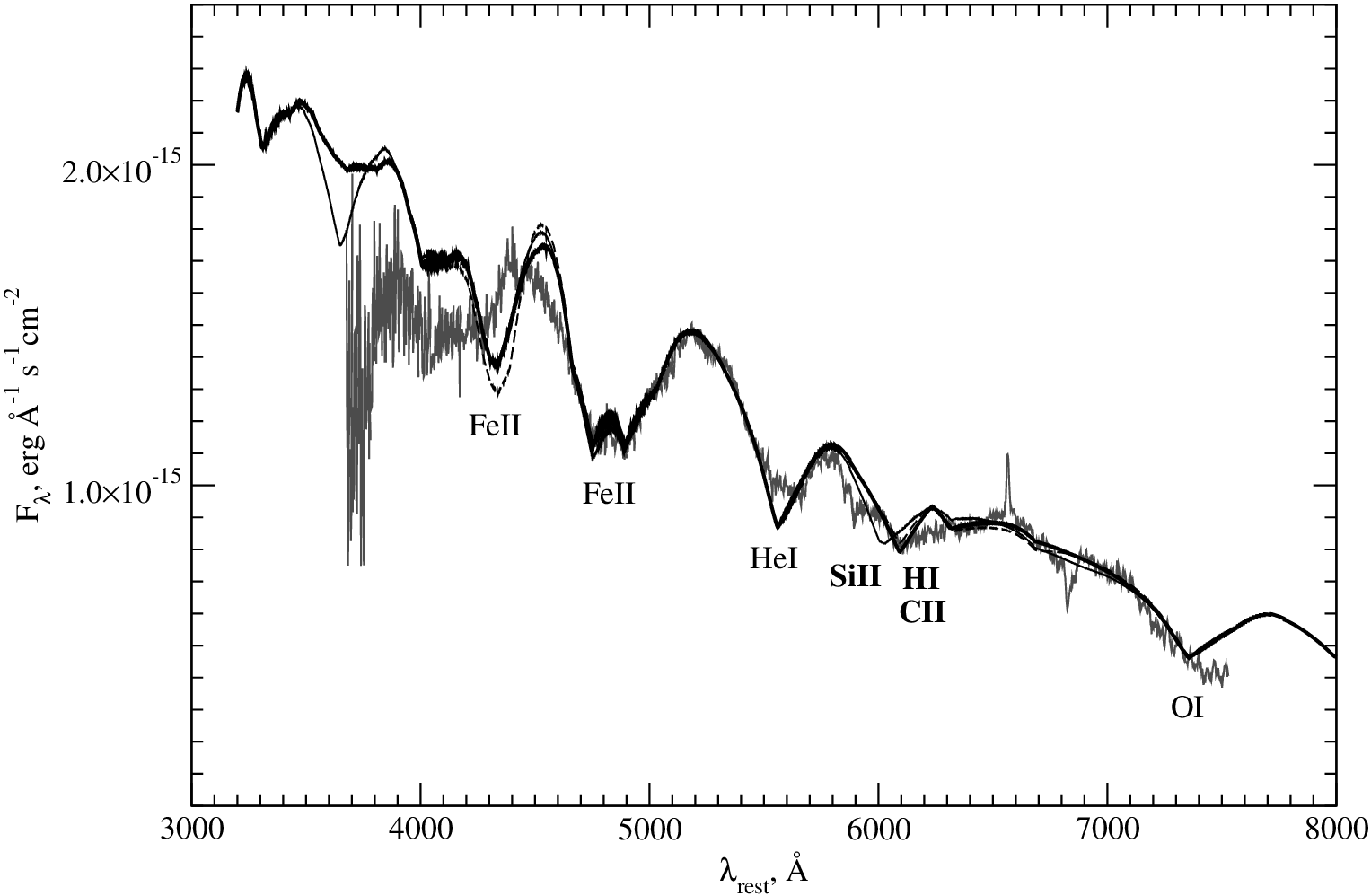}
   \captionstyle{normal}
\caption{ The spectrum of the supernova SN\,2008D obtained with
the BTA on January 16, 2008 in 6.48 days after the explosion (in
the reference frame linked with the object, $z=0$) is shown by the
thin grey noisy line; the thick black line marks the model
spectrum which contains H\,I; the model spectrum containing C\,II
instead of hydrogen is marked by the thin dashed line; the model
spectrum with Si\,II instead of hydrogen is marked by the solid
thin black line.}

\label{6.48:Moskvitin_n}
\end{figure*}

\begin{figure*}
\setcaptionmargin{5mm}
\onelinecaptionsfalse
    \includegraphics[width=15cm, bb=20 10 800 500, clip]{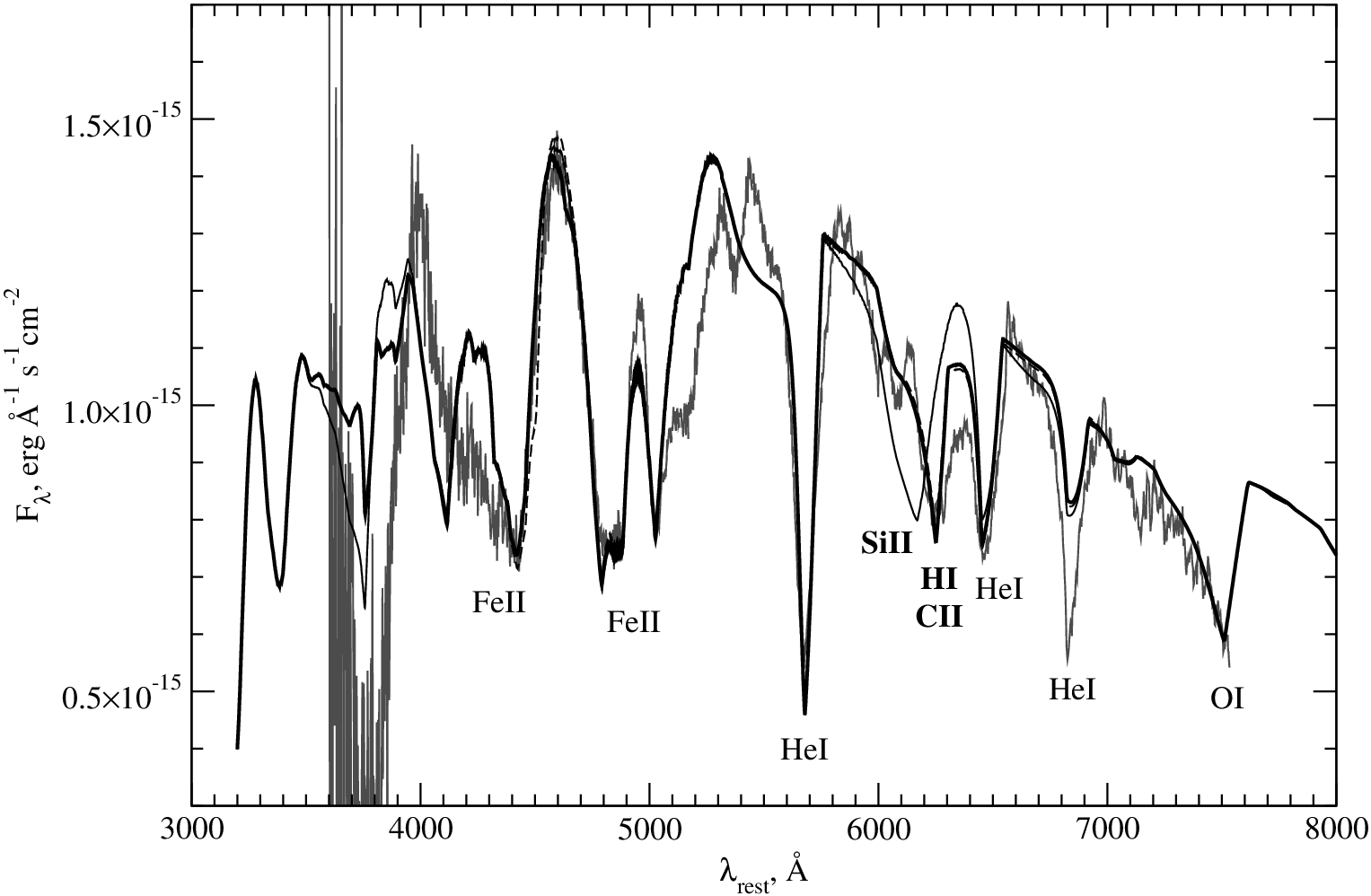}
   \captionstyle{normal}
\caption{The spectrum of the supernova SN\,2008D obtained with the
BTA on February 6, 2008 in 27.62 days after the explosion. The
markings are the same as in Fig. \ref{6.48:Moskvitin_n}.}
\label{27.62:Moskvitin_n}
\end{figure*}

The main absorption features are noticeable in both spectra (see
Figs. \ref{6.48:Moskvitin_n} and \ref{27.62:Moskvitin_n}) as P Cyg
profiles of He\,I, Fe\,II, O\,I  and, presumably, H\,I (a spectral
feature near 6200\,\AA{}) lines. The main models containing H\,I
as a candidate describing the above-mentioned feature are shown by
the thick black line, and the observed spectra---by the thin grey
noisy line.

%Parameters of main models are given in Tables \ref{2008D_6.48} and
%\ref{2008D_27.62}. The photosphere velocity was taken equal to the
%velocity of lines of ionized iron.

The first spectrum is best described by the model with the
following parameters: photosphere temperature \mbox{$T_{bb}=8700$
K}, velocity of the photosphere  \linebreak $V_{phot}=17000$ km/s,
maximum matter expansion velocity in the envelopes \mbox{$V_{max}
= 70000$ km/s}. The parameters of the best model for the second
spectrum are as follows: \mbox{$T_{bb}=6000$ K},
\mbox{$V_{phot}=8500$ km/s}, \mbox{$V_{max}=70000$ km/s}.

The temperature of the photosphere was determined by the slope of
the observed spectrum. Since the shape of the spectra is very
different from the Planck curve, the ions, describing the main
spectral features were immediately included in the calculation.
The photosphere velocity was adopted as equal to the velocity of
the ionized iron \mbox{lines \cite{Elmhamdi2006:Moskvitin_n}.} The
maximal velocity at the upper boundary of all envelopes $V_{max}$
was estimated from the blue part of the P Cyg line profiles.
Optical depth of the reference line
\cite{Elmhamdi2006:Moskvitin_n} was selected in a way to best
describe the absorption components of the P Cyg profile. The
\texttt{SYNOW} code allows applying one of the two laws of the
matter density decrease with distance: power \linebreak ($\tau(r)
\propto v(r)^{-n}$, where $v \sim r$), and exponential \linebreak
($\tau(r) \propto \exp(-v(r)/V_e)$, where $V_e$ is the
characteristic velocity of the layer, which determines the size of
the envelope and the width of the absorptions in the observed
spectrum).  We used the exponential law in our calculations. The
parameters of the main models that take hydrogen into account are
presented in Table \ref{2008D_6.48_27.62:Moskvitin_n}.

\begin{table*}
\setcaptionmargin{0mm} \onelinecaptionstrue \captionstyle{normal}
\caption{Parameters of the models which best describe both spectra
of SN\,2008D. The parameters correspond to the models, marked with
thick black lines in Figs. \ref{6.48:Moskvitin_n} and
\ref{27.62:Moskvitin_n}; $\tau$ is the optical depth of the
reference line of every ion; $V_{min}$ is the minimal layer
velocity of every ion; $V_e$ is the characteristic velocity in the
applied relation between the optical depth and the expansion
velocity $\tau(r) \sim \exp(-v(r)/V_e)$, where $v \sim r$;
$V_{phot}$ is the photosphere velocity; $V_{max}$ is the velocity
at the upper boundary of all envelopes; $T_{bb}$ is the black body
temperature of the photosphere, forming a continuous spectrum.}
\label{2008D_6.48_27.62:Moskvitin_n}
\bigskip
\begin{tabular}{l|c|c|c|c|c|c|c|c}

\hline
%==============================================================================================================================
Parameters                 & \multicolumn{4}{c|}{First spectrum (6.48 days)} & \multicolumn{4}{c}{Second spectrum (27.62 days)}\\
\hline

Ions                      & H\,I  & He\,I & Fe\,II & O\,I  & H\,I  & He\,I & Fe\,II & O\,I  \\
\hline
$\tau$                    &  0.4  & 0.5   &  1.0   &  0.5  & 0.5   & 3.0   & 2.0    & 0.3   \\
\hline
$V_{min}$, km/s   & 23000 & 17000 &  17000 & 17000 & 15000 & 10500 & 8500   & 10500 \\
\hline
$V_e$, km/s       & 8000  & 4000  &  3000  &  4000 & 3000  & 1000  & 3000   & 4000  \\
\hline
$V_{phot}$, km/s  & \multicolumn{4}{c|}{17000}    & \multicolumn{4}{c}{8500}\\
\hline
$V_{max}$, km/s   & \multicolumn{4}{c|}{70000}    & \multicolumn{4}{c}{70000}\\
\hline
$T_{bb}$, K            & \multicolumn{4}{c|}{8700}     & \multicolumn{4}{c}{6000}\\
%==============================================================================================================================
\hline
\end{tabular}
\end{table*}

As an alternative interpretation of the spectral feature near
6200\,\AA{}, we considered the model calculations of synthetic
spectra containing the lines of Si\,II and C\,II instead of the
H\,I lines. These ions were frequently mentioned in the literature
on modeling the type Ib-c supernovae spectra
\cite{Elmhamdi2006:Moskvitin_n, Valenti2007:Moskvitin_n}. In Figs.
\ref{6.48:Moskvitin_n} and \ref{27.62:Moskvitin_n} the thin solid
lines correspond to the models with silicon taken instead of
hydrogen: \mbox{$\tau=0.005$}, \mbox{$V_{min}=17000$ km/s},
\mbox{$V_e=10000$  km/s}  for the first spectrum (all the other
parameters are identical to those in the model with hydrogen), and
\mbox{$\tau=0.5$}, \mbox{$V_{min}=8500$ km/s}, \mbox{$V_e=5000$
km/s} for the second spectrum. The models containing the Si\,II
lines \linebreak ($\lambda_{rest}$ = 6347\,\AA) instead of H\,I
lines are velocity-limited by the photosphere and can not describe
the observed spectral features near 6200\,\AA{} according to the
modern concepts on the radial stratification of elements
\cite{Elmhamdi2006:Moskvitin_n}.

The absorption feature near 6200\,\AA{} can as well be explained
by the presence of ionized carbon C\,II with the following model
parameters: \mbox{$\tau=0.005$,} \mbox{$V_{min}=24000$ km/s,}
\mbox{$V_e=10000$ km/s} for the first spectrum, and
\mbox{$\tau=0.0008$}, \mbox{$V_{min}=16000$ km/s}, \linebreak
$V_e=3000$ km/s for the second spectrum (the thin dashed lines in
Figs. \ref{6.48:Moskvitin_n} and \ref{27.62:Moskvitin_n}).

Thus, Si\,II can be excluded from the calculations of these two
spectra of SN\,2008D. C\,II line \linebreak
($\lambda_{rest}=6580$\,\AA) remains a possible alternative for
the description of the absorption feature near 6200\,\AA. The
above mentioned feature can possibly be formed by a blend of the
H$\alpha$ and C\,II lines. It is obvious that the final conclusion
requires observations with a higher spectral resolution.

\subsection{Modeling of SN\,2006Aj Spectra}

The spectra of SN\,2006aj obtained with the BTA in 2.55 and 3.55
days after the gamma-ray burst were modeled with the
\texttt{SYNOW} code. The first observed spectrum is best modeled
with the photosphere expansion velocity of $33000$ km/s. This
parameter is within measurement errors of the photosphere
expansion velocity, measured from the \textit{Swift/XRT/UVOT} data
((2.7$\pm$0.8)$\times$10$^{4}$ km/s
\cite{Blustin2007:Moskvitin_n}). At the moment we started our
spectral observations (roughly straight after the last observation
of the UV flash with the \textit{Swift/UVOT}), the photosphere
velocity remained within approximately the same limits. Moreover,
the early spectrum shows a broad continuum depression at
\mbox{5900--6300\,\AA{}} and an almost imperceptible excess at
6300--6900\,\AA, which is best described by the H$\alpha$ P Cyg
profile with the same velocity of \linebreak $33000$ km/s. For the
fitting we used different models of the case where the layers are
undetached from the photosphere, but the velocities of the
photosphere and ions remained equal to $33000$ km/s. The model
parameters, the fitting of the observed spectrum, and detailed
comments are given in \cite{Sonbas2008:Moskvitin_n}.

The velocities best describing the second spectrum (3.55 days
after the burst) are within the limits 18000 km/s $\leqslant
\{V_{phot},\,V_{min}\}\leqslant 24000$ km/s. The parameters of the
model spectrum describing the observed spectrum, and the selected
values of model parameters are laid out in
~\cite{Sonbas2008:Moskvitin_n}.

\begin{figure*}
\setcaptionmargin{5mm}
\onelinecaptionsfalse
   \centerline{\includegraphics[width=15cm, bb=40 10 750 500, clip]{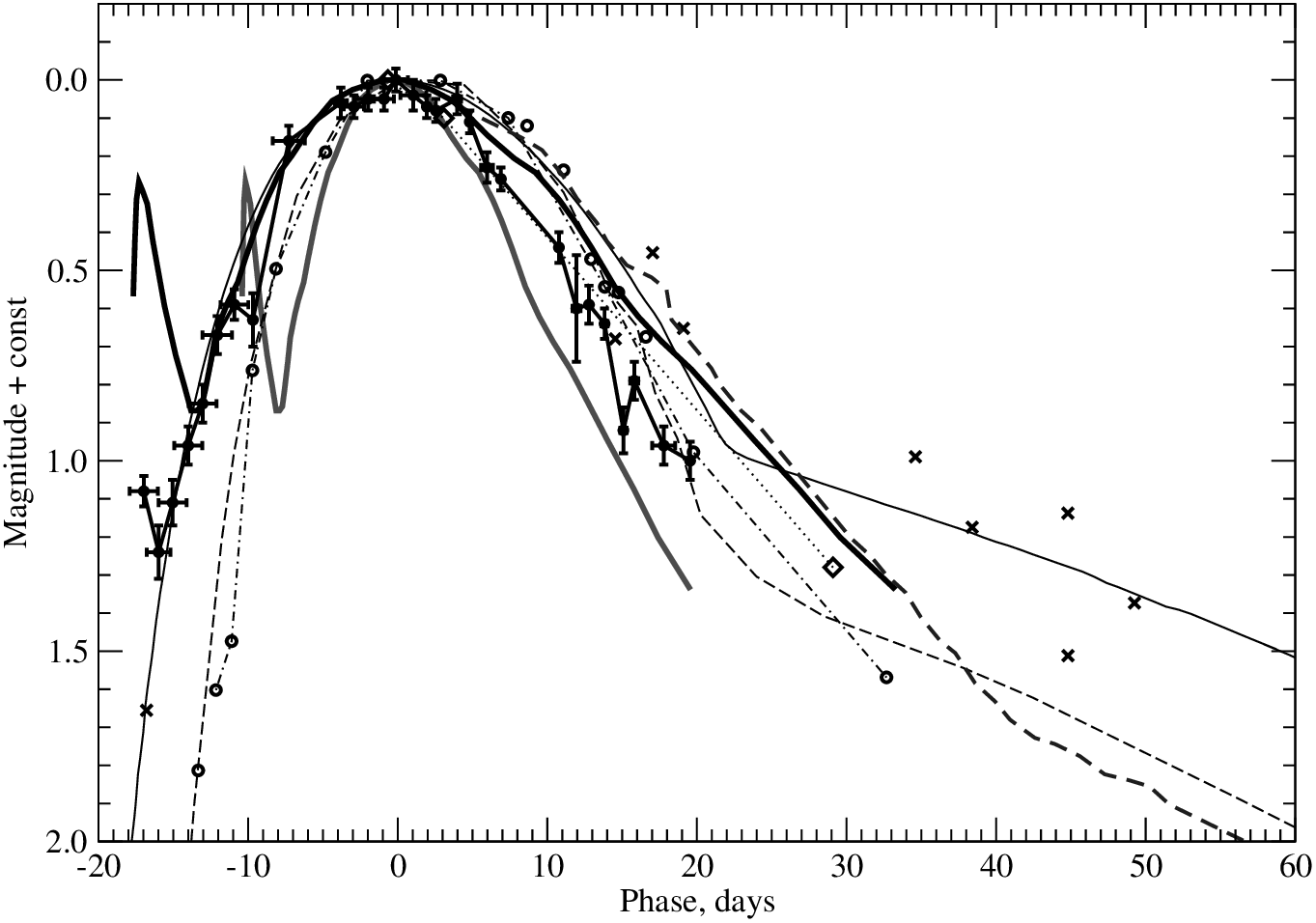}}
      \captionstyle{normal}
\caption{Comparison of light curves of the following supernovae in
the V-band: SN\,1983N \cite{Shlegel_Kirshner1989:Moskvitin_n,
Newmark1992:Moskvitin_n, Baron1993:Moskvitin_n}; SN\,1984L
\cite{Baron1993:Moskvitin_n};
SN\,1999dn~\cite{Benetti2002:Moskvitin_n}; SN\,2000H
\cite{Richardson2006:Moskvitin_n}; SN\,2008D
\cite{Modjaz2008:Moskvitin_n}; SN\,2006aj
\cite{Sonbas2008:Moskvitin_n}. The thick grey line shows the light
curve of  SN\,2006aj without scaling along the time axis; the
thick black line shows the same light curve scaled along the time
axis with a factor of 1.7. The SN\,2008D light curve is  shown
with the black line connecting the points with bars.}
\label{LCs_V:Moskvitin_n}
\end{figure*}

Using the best models of the SN\,2006aj and SN\,2008D spectra, we
measured the hydrogen and helium envelope velocities, and the
velocities of photosphere expansion for the respective observation
epochs (see Table \ref{table_vel:Moskvitin_n}). The earliest
spectra of SN\,2006aj demonstrate one the highest observed
expansion velocities of the order of 0.1 of the speed of light.
The teams of Kunugasa and Sahu (\cite{Kunugasa2002:Moskvitin_n,
Sahu2009:Moskvitin_n}) describe similar expansion velocities
observed in the early spectra of type Ic supernovae SN\,2002ap and
SN\,2007ru, which indicates  the presence of not just a unique
object but rather a class of events capable of showing such high
velocities.

\begin{table*}
% \begin{center}
 \setcaptionmargin{5mm} \onelinecaptionsfalse
\captionstyle{normal} \caption{ Velocity variation of the hydrogen
and helium envelopes and photosphere expansion of the supernovae
SN\,2006aj and SN\,2008D after the explosion. The values in
brackets indicate time after the maximum. The time of the maximum
for SN\,2006aj was determined as 10.4 days after the explosion
(based on the V-band light \mbox{curve)
\cite{Sollerman2006:Moskvitin_n},} and 19 days for SN\,2008D
\cite{Mazzali2008:Moskvitin_n}.} \label{table_vel:Moskvitin_n}
  \bigskip
  \begin{tabular}{l|l|l|l|l}
   \hline
\multirow{2}{*}{ }              & \multicolumn{2}{c|}{SN~2006aj} & \multicolumn{2}{c}{SN~2008D}\\
   \cline{2-5}
                                 & 2.55 $(-7.85)$ & 3.55 $(-6.85)$ & $6.48 (-12)$ & 27.62 (+8) \\
   \hline

$V_{photosphere}$ , km/s  & 33000        & 18000        & 17000       & 8500        \\
$V_{hydrogen}$ , km/s      & 33000        & 24000        & 23000       & 15000       \\
$V_{helium}$ , km/s        & 33000        & 24000        & 17000       & 10500       \\
   \hline
  \end{tabular}
% \end{center}
\end{table*}

\begin{figure*}
\setcaptionmargin{5mm}
\onelinecaptionsfalse
\includegraphics[width=15cm, bb=30 00 750 500, clip]{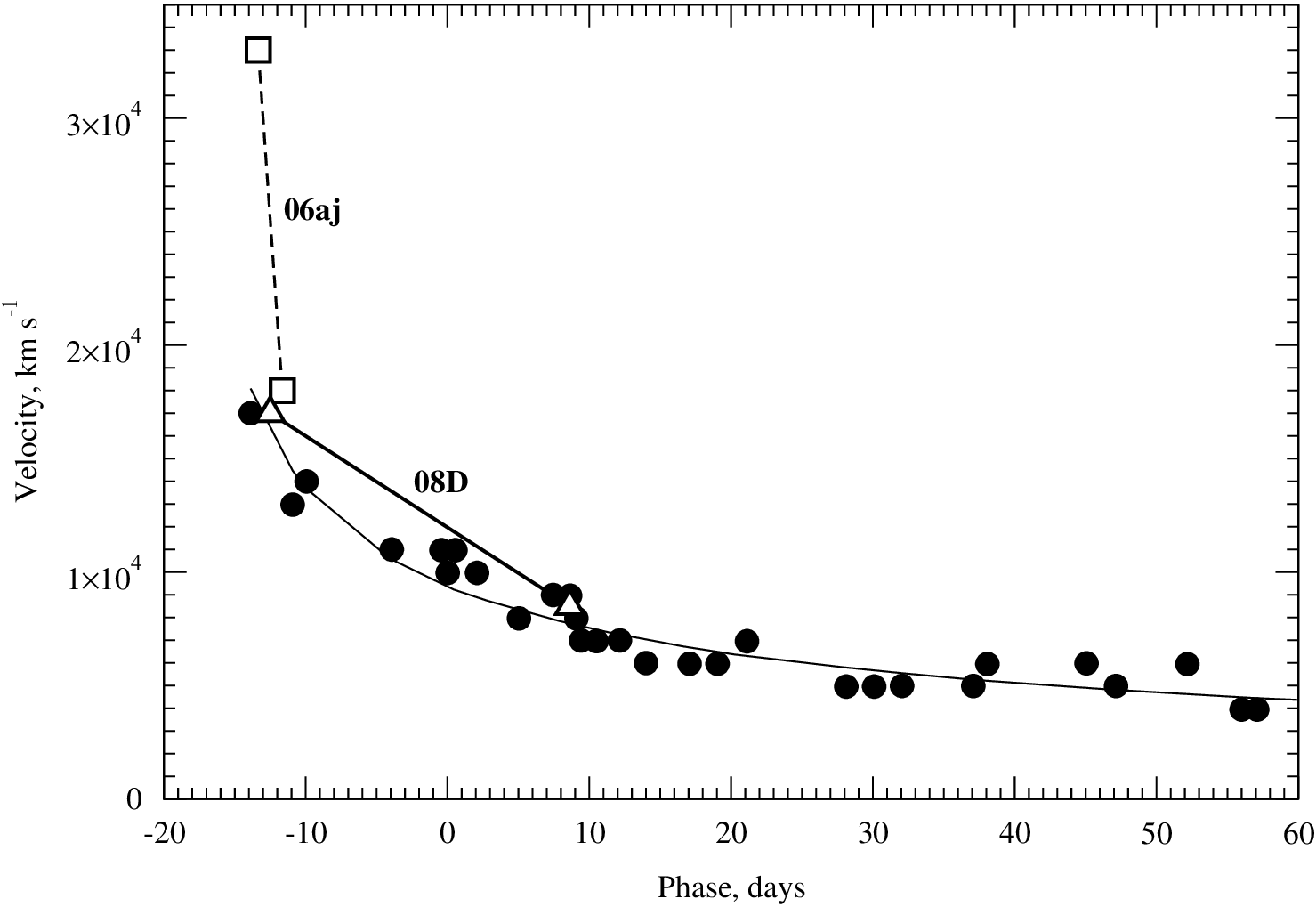}
      \captionstyle{normal}
\caption{Variation of photosphere velocities, measured from the Fe
II lines \cite{Branch2002:Moskvitin_n}. Photosphere velocities
measured from our spectra are denoted as  triangles for SN\,2008D
and squares for SN\,2006aj. The scaling factor 1.7 was applied to
the SN\,2006aj photosphere velocity.} \label{fig22:Moskvitin_n}
\end{figure*}

\begin{figure*}
\setcaptionmargin{5mm}
\onelinecaptionsfalse
   \includegraphics[width=15cm, bb=30 0 750 500, clip]{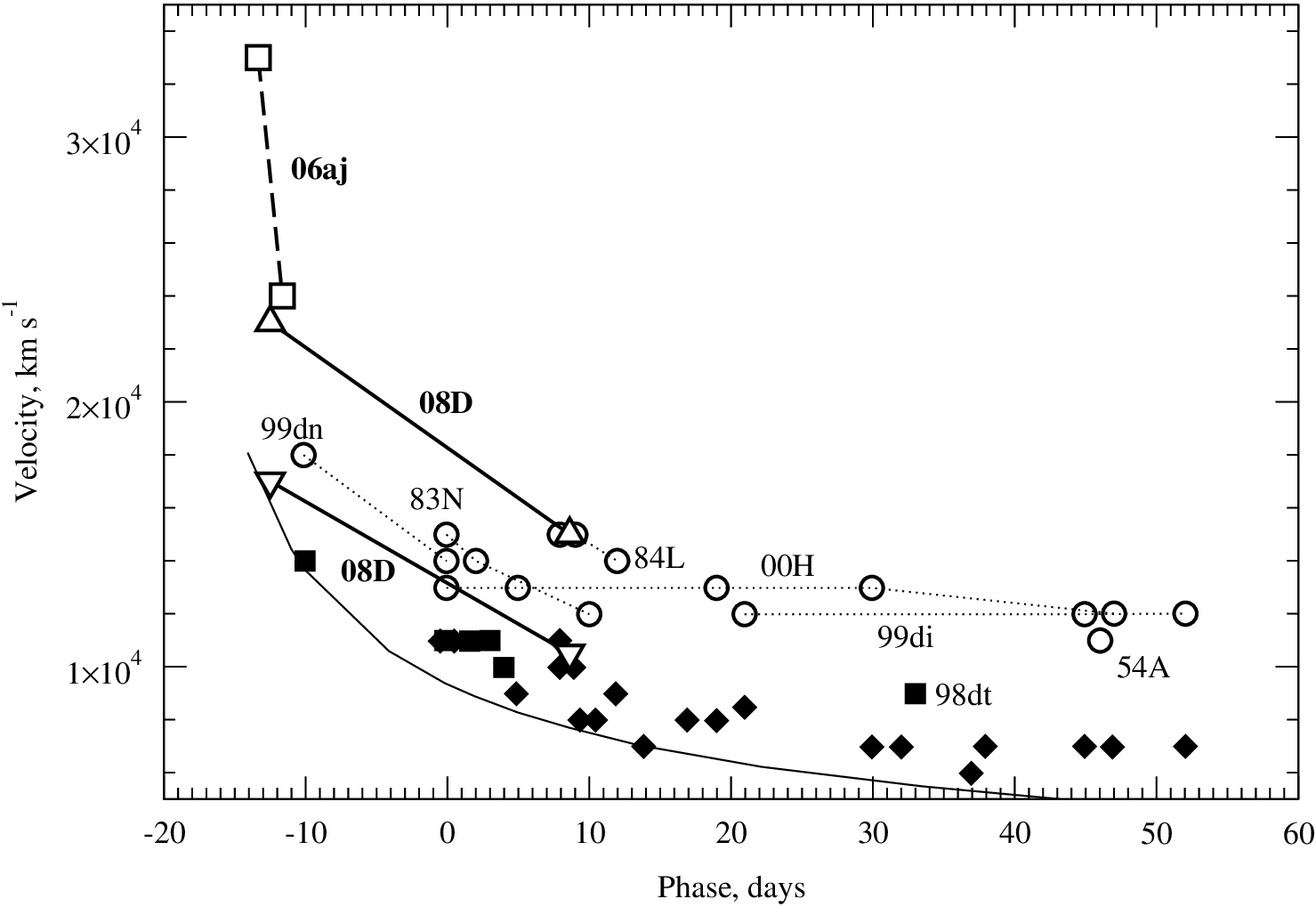}
\captionstyle{normal} \caption{ Minimal velocities from He\,I
lines (filled squares for the case of layers undetached from the
photosphere, filled diamonds for the detached lines), and H\,I
lines (empty circles, always in detached layers)
\cite{Branch2002:Moskvitin_n}. The curve represents the
photosphere velocity drop power law from Fig.
\ref{fig22:Moskvitin_n}. Minimal velocities of He\,I lines for
SN\,2008D are indicated by triangles up, H\,I line velocities---by
triangles down. The velocities of hydrogen and helium lines for
SN\,2006aj are indicated by squares. The scaling factor 1.7 was
applied to SN\,2006aj envelope velocities.}
\label{fig23:Moskvitin_n}
\end{figure*}

We compared the velocities obtained for SN\,2008D and SN\,20006aj
with an empiric law of photosphere and envelope velocity decrease
for 11 type Ib supernovae from \cite{Branch2002:Moskvitin_n} using
the scaling factor for SN\,2006aj that was obtained based on a
comparison of the light curves of the supernovae sample. This
comparison is presented in Figs. \ref{fig22:Moskvitin_n} and
\ref{fig23:Moskvitin_n}.

%___________________________fig22

%___________________________fig23

\section{PHOTOSPHERE AND ENVELOPE VELOCITIES}\label{Vel_phot_env:Moskvitin_n}

Light curves of different type Ib, Ic and Ib-c supernovae mainly
differ in absolute fluxes, widths of bell-shaped peaks near the
main maxima, and their behavior at late phases of the flash.
However, the light curve shapes of different supernovae near the
maximum are similar to each other. Valenti et al.
\cite{Valenti2007:Moskvitin_n}  compared the light curves of SNe
of different types via scaling along the time axis. We applied an
identical method when comparing the light curves of SN\,2008D and
SN\,2006aj in different bands with the light curves of SNe from
\cite{Branch2002:Moskvitin_n}. In the case of SN\,2006aj, we
applied the scaling factor 1.7 (see Fig. \ref{LCs_V:Moskvitin_n}).
A comparison of light curves of the SNe not observed in the
\mbox{V-band} with the data available in the B- and R-bands
confirmed the accuracy of our choice of the scaling factor for the
SN\,2006aj light curve.

\section{CONCLUSIONS}\label{results:Moskvitin_n}

An agreement between the velocities of photosphere expansion and
those of the helium/hydrogen envelopes of the SNe associated with
the X-ray flash XRF\,080109 and the gamma-ray burst GBR\,060218
argues for similar explosion dynamics of gamma-ray bursts and
X-ray flashes.

A comparison with analogous velocities of classical type Ib
supernovae not accompanied by gamma- or X-ray bursts may indicate
common properties of the progenitors of these two classes of
phenomena, and similar explosion mechanisms.

Previously, an idea about the presence of hydrogen in the spectra
of type Ib and Ic supernovae was suggested
~\cite{Branch2006:Moskvitin_n, Elmhamdi2006:Moskvitin_n}. The idea
was confirmed in our paper ~\cite{Sonbas2008:Moskvitin_n}
dedicated to the study of early spectra of SN\,2006aj.

Branch et al. \cite{Branch2006:Moskvitin_n} list the main
candidates for the description of a spectral feature near
6200\,\AA{} observed in the early spectra of supernovae. Apart
from the hydrogen line H$\alpha~ \lambda6563$ these are: Si\,II
$\lambda6355$ spaced from H$\alpha$ at 9500 km/s, Ne\,I
$\lambda6402$ at 7360 km/s, and only C\,II $\lambda6580$ is spaced
from the hydrogen line at 777 km/s, which can be within velocity
measurement errors. Candidate velocities must agree with the
photosphere velocity determined from the heaviest ions (e.g., from
the ionized iron). Reasoning from the results of modeling the
SN\,2006aj and SN\,2008D spectra, it can be concluded that the
most possible sources of absorption near $\lambda$6200 are
H$\alpha$~ $\lambda6563$ and C\,II $\lambda6580$. To choose
between these candidates we compared the velocity variability of
hydrogen envelopes of the two SNe under study with the classical
supernovae. The results of comparison of the velocity evolutions
of different supernovae argue for the existence of
hydrogen-containing envelopes around massive progenitor stars that
appear before the explosion due to more or less powerful stellar
winds.

Detection of P Cyg profile absorption components of other Balmer
lines, in particular H$\beta$ $\lambda4861$, can become a decisive
argument in favor of hydrogen presence in the envelope. Since the
optical depth of H$\beta$ line is smaller than that of H$\alpha$,
there arises a problem of its detection in a noisy spectrum among
numerous lines of heavy metals, particularly, Fe\,II. Detailed
long-term observations of the evolution of subtle details in the
spectra obtained with a high signal/noise ratio can solve this
problem.

% Nevertheless, the identi?cation of the 6200 A absorption is plagued by ambiguities. The
% strongest line of Ne I, ?6402, is about 2200 km s?1 to the red of Si II ?6355, and H-alpha
% ?6563 is only about 800 km s?1 to the blue of C II ?6580. Because these four ions, each of
% which could appear in supernova spectra (although Ne I probably would require nonthermal
% excitation), have their strongest optical lines to the red, but not too far to the red, of the

\begin{acknowledgements}

The work has been supported by the RNP grant no. 2.1.1.3483 of the
Federal Agency for Education of the Russian Federation. The
authors are grateful to Evgeny Semenko and Azamat Valeev for their
help in the preparation of this paper.

\end{acknowledgements}


\begin{thebibliography}{}

\bibitem{Chevalier_Fransson2008:Moskvitin_n}
R. A. Chevalier and C. Fransson, \apj{} \textbf{683}, L135 (2008).

\bibitem{Soderberg2008:Moskvitin_n}
A. M. Soderberg et al., Nature \textbf{453}, 469 (2008).

\bibitem{Modjaz2008:Moskvitin_n}
M. Modjaz et al., \apj{} \textbf{702}, 226 (2009).

\bibitem{Blustin2007:Moskvitin_n}
A. J. Blustin, Roy. Soc. Phil. Trans. \textbf{A 365}, 1263 (2007).

\bibitem{Campana2006:Moskvitin_n}
S. Campana et al., Nature \textbf{442}, 1008 (2006).

\bibitem{Sonbas2008:Moskvitin_n}
E. Sonbas et al., \ab{} \textbf{63}, 228 (2008).

\bibitem{Schlegel1998:Moskvitin_n}
D. J. Schlegel, D. P. Finkbeiner, and M. Davis, \apj{} \textbf{500}, 525 (1998).
%Schlegel, David J.; Finkbeiner, Douglas P.; Davis, Marc


\bibitem{Cardelli1989:Moskvitin_n}
J. A. Cardelli, G. C. Clayton, and J. S. Mathis, \apj{} \textbf{345}, 245 (1989).
%Cardelli, Jason A.; Clayton, Geoffrey C.; Mathis, John S.

\bibitem{Sollerman2006:Moskvitin_n}
J. Sollerman et al., A\&A \textbf{454}, 503 (2006).

\bibitem{Branch2001:Moskvitin_n}
D. Branch, E. Baron, and D. J. Jeffery, \textit{Supernovae and Gamma-Ray Bursts}, Ed. by K. W. Weiler, Lecture Notes in Physics
      \textbf{598}, 47 (2001).

\bibitem{Branch2002:Moskvitin_n}
D. Branch et al., \apj{} \textbf{566}, 1005 (2002).

\bibitem{Baron2005:Moskvitin_n}
E. Baron, P. E. Nugent, D. Branch, and P. H. Hauschildt, ASP Conference Series \textbf{342}, 351 (2005).


\bibitem{Elmhamdi2006:Moskvitin_n}
A. Elmhamdi et al., A\&A \textbf{450}, 305 (2006).

\bibitem{Valenti2007:Moskvitin_n}
S. Valenti et al.,
\mnras{} \textbf{383}, 1485 (2007).
% S.Valenti et al., ?The Broad-lined Type Ic SN 2003jd?
%  arXiv:0710.5173v1 [astro-ph] 26 Oct 2007.
% Monthly Notices of the Royal Astronomical Society, Volume 383, Issue 4, pp. 1485-1500

%****************************LCs_V

\bibitem{Shlegel_Kirshner1989:Moskvitin_n}
E. M. Shlegel and R. P. Kirshner, \aj{} \textbf{98}, 577 (1989).
%, ?The type Ib supernova 1984L in NGC 991?

\bibitem{Newmark1992:Moskvitin_n}
J. S. Newmark et al.
``IUE observing guide'', IUE Observatory and Computer Sciences Corporation, May 1992, paragraph 2.3,
  \url{http://archive.stsci.edu/iue/instrument/obs_guide/node8.html}

\bibitem{Baron1993:Moskvitin_n}
E. Baron, T. R. Young, and D. Branch, \apj{} \textbf{409}, 417 (1993).

\bibitem{Benetti2002:Moskvitin_n}
S. Benetti et al., \mnras{} \textbf{336}, 91 (2002).

%?The Exceptionally Bright Type Ib Supernova 1991D?. B

\bibitem{Richardson2006:Moskvitin_n}
D. Richardson, D. Branch, and E. Baron, \aj{} \textbf{131}, 2233 (2006).
%. ?Absolute-Magnitude Distributions and Light Curves of Stripped-Envelope Supernovae?

%****************************LCs_V
\bibitem{Mazzali2008:Moskvitin_n}
 P. A. Mazzali et al., Science \textbf{321}, 1185 (2008).


\bibitem{Kunugasa2002:Moskvitin_n}
K. Kunugasa et al., \apj{} \textbf{577}, L97 (2002).

\bibitem{Sahu2009:Moskvitin_n}

D. K. Sahu et al., \apj{} \textbf{697}, 676 (2009).


\bibitem{Branch2006:Moskvitin_n}
D. Branch et al.,
\pasp{} \textbf{118}, 791 (2006).

%_________________________________________________

\end{thebibliography}
\end{document}